\pdfoutput=1
\documentclass[9pt,twocolumn,letterpaper]{extarticle}

\usepackage[letterpaper,textwidth=42pc,textheight=636pt,columnsep=2pc,top=0.8in,hcentering,headheight=11pt,headsep=12pt]{geometry}
\usepackage[T1]{fontenc}
\usepackage{times}
\usepackage{graphicx}
\usepackage{amsmath,amssymb}
\usepackage{etoolbox}
\usepackage[font=small,labelfont=bf]{caption}
\usepackage{cite}
\usepackage{xcolor}
\usepackage[hyperfootnotes=false]{hyperref}
\definecolor{linkblue}{rgb}{0.1,0.2,0.55}
\hypersetup{
  colorlinks=true, allcolors=linkblue,
  pdftitle={emg2face: Expressive Facial Animation with High-Density Surface EMG},
  pdfauthor={Ganidhu Abey, Wendy Greening, Ashika Kamboj, Leonhard Helminger,
             Abhijeet Ghosh, Karel Petranek, Sergio Orts Escolano, Dinesh K. Pai}}

\makeatletter

\renewcommand\section{\@startsection{section}{1}{\z@}%
  {-2.2ex \@plus -1ex \@minus -.2ex}{1.2ex \@plus .2ex}{\normalfont\large\bfseries}}
\renewcommand\subsection{\@startsection{subsection}{2}{\z@}%
  {-1.8ex \@plus -1ex \@minus -.2ex}{0.8ex \@plus .2ex}{\normalfont\normalsize\bfseries}}

\newcommand{\@teaser}{}
\newcommand{\teaser}[1]{\gdef\@teaser{#1}}

\newcommand{\arxivauthors}{%
  Ganidhu Abey$^{1*}$\quad
  Wendy Greening$^{1*}$\quad
  Ashika Kamboj$^{1*}$\quad
  Leonhard Helminger$^{2}$\\[3pt]
  Abhijeet Ghosh$^{2,3}$\quad
  Karel Petranek$^{2}$\quad
  Sergio Orts Escolano$^{2}$\quad
  Dinesh K. Pai$^{1,4\dagger}$}
\newcommand{\arxivaffiliations}{%
  $^1$University of British Columbia\qquad
  $^2$Google\qquad
  $^3$Imperial College London\qquad
  $^4$Vital Mechanics Research}

\renewcommand{\maketitle}{%
  \twocolumn[{%
    \centering
    {\huge\bfseries\@title\par}\vspace{14pt}
    {\large\arxivauthors\par}\vspace{6pt}
    {\normalsize\arxivaffiliations\par}\vspace{14pt}
    \begin{minipage}{\textwidth}\captionsetup{type=figure}\@teaser\end{minipage}\par
    \vspace{16pt}}]%
  \begingroup
    \renewcommand{\thefootnote}{}%
    \footnotetext{$^*$Equal contributions, listed alphabetically.}%
    \footnotetext{$^\dagger$Corresponding author: \href{mailto:pai@cs.ubc.ca}{pai@cs.ubc.ca}}%
  \endgroup
  \thispagestyle{plain}}

\AtBeginEnvironment{thebibliography}{\small}

\def\ps@arxiv{%
  \def\@oddhead{\hfil{\small\itshape Abey et al.\ / emg2face}\hfil}%
  \let\@evenhead\@oddhead
  \def\@oddfoot{\hfil\thepage\hfil}%
  \let\@evenfoot\@oddfoot}

\makeatother

\title{emg2face: \\ Expressive Facial Animation with High-Density Surface EMG}

\graphicspath{{figures/}}
\newlength{\setupfigh}

\begin{document}

\teaser{
 \centering
 \setlength{\setupfigh}{0.225\textwidth}%
 \includegraphics[height=\setupfigh]{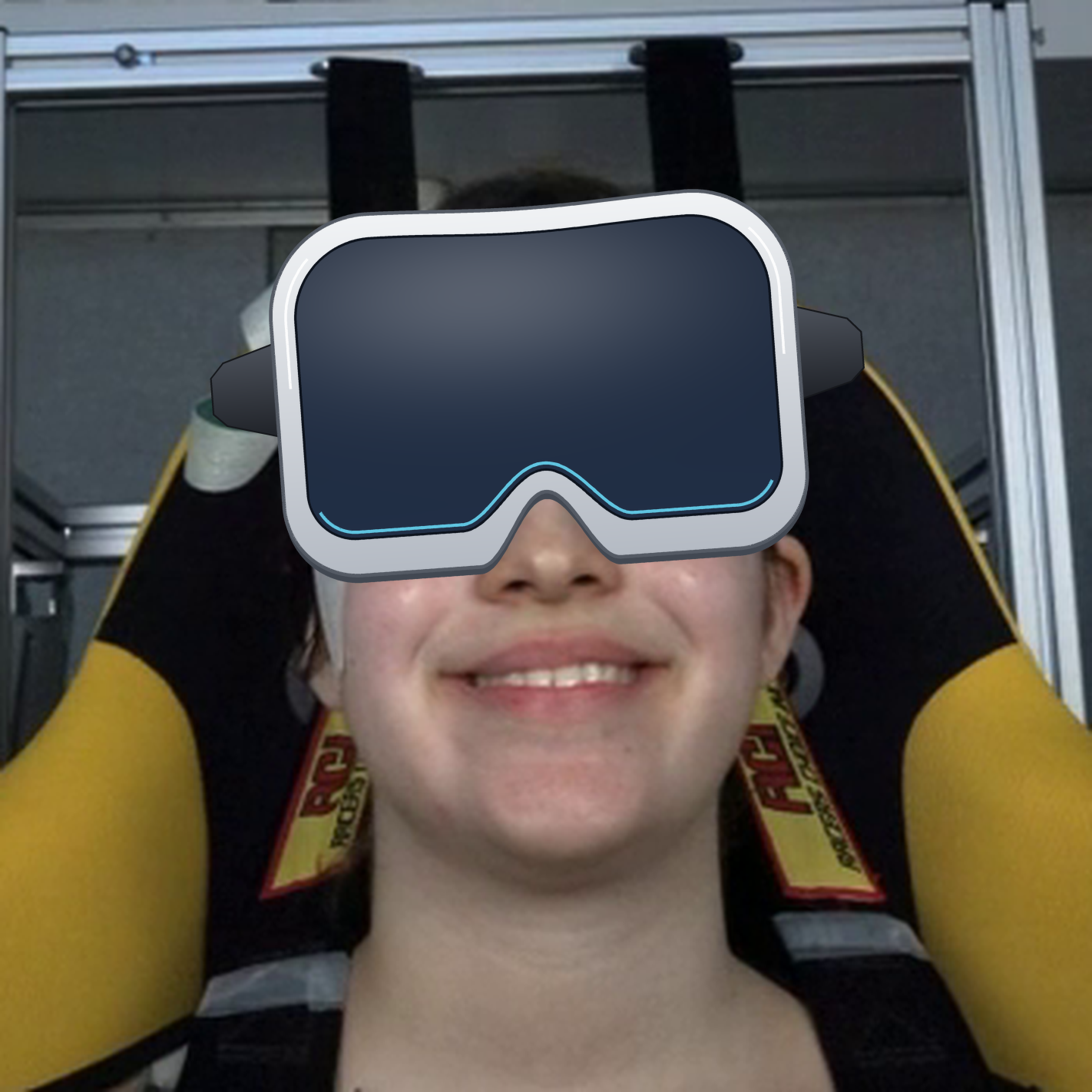}\hfill
 \raisebox{0.42\setupfigh}{\includegraphics[height=0.16\setupfigh]{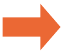}}\hfill
 \includegraphics[height=\setupfigh]{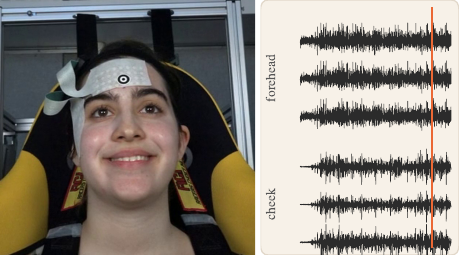}\hfill
 \raisebox{0.42\setupfigh}{\includegraphics[height=0.16\setupfigh]{teaser_arrow}}\hfill
 \includegraphics[height=\setupfigh]{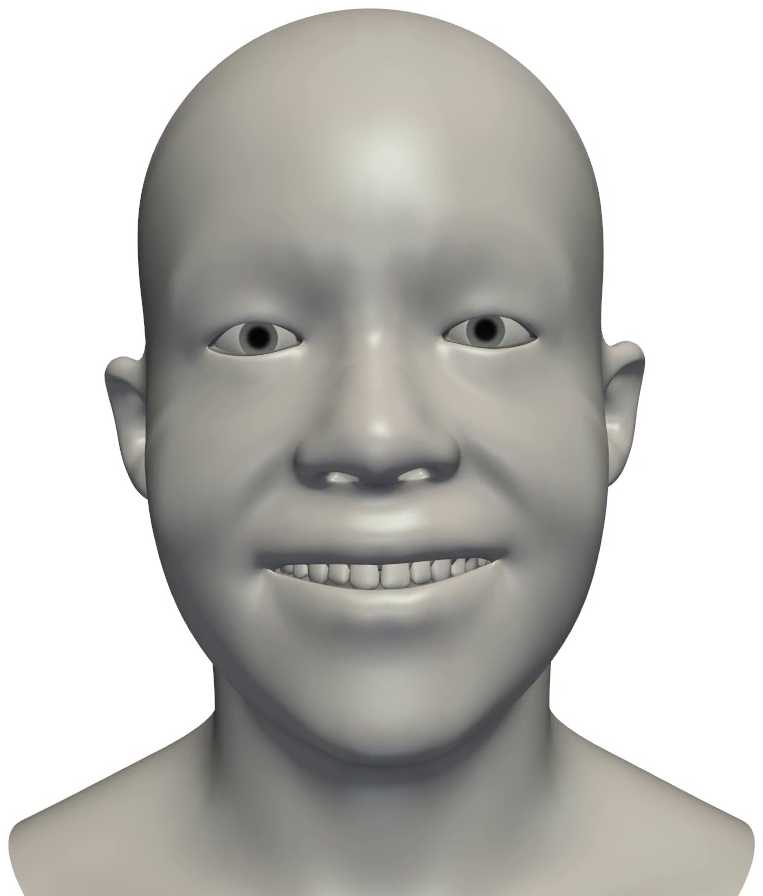}
  \caption{We decode facial expressions from facial muscle activity, even when a head-mounted display (drawn here for illustration) hides the upper face from optical capture (left). Two textile electrode grids on the forehead and the cheek record HD-sEMG (middle) from which a network predicts the expression of the high-resolution GNM head model (right).}
\label{fig:teaser}
}

\maketitle
\begin{abstract}
   Facial movements convey subtle and important information that is critical for human social communication. Optical methods for face capture are difficult or impossible to use when the face is occluded by head-mounted devices (HMDs), such as VR headsets. Even with a clear line of sight to the face, such methods raise privacy concerns and require head-mounted capture rigs that offset cameras and lighting from the face. We show that high-density surface electromyography (HD-sEMG) provides a viable non-optical alternative that addresses these challenges.

We measured 64 EMG channels, using two textile EMG grids, with 32 from the forehead (typically occluded by an HMD) and 32 from the side of the face. EMG data were digitized at 2048 Hz and filtered. Facial movements were simultaneously recorded and used to estimate 478 3D facial landmarks using MediaPipe's Face Landmarker. A major challenge in such multimodal recordings is synchronizing EMG and video data, which have different sampling frequencies and independent clocks. We developed a novel synchronization method using analog audio bursts that is capable of sub-millisecond synchronization. We also developed a staged fitting method that fits a recent high-resolution parametric head model (GNM), with 253 identity blendshapes and 383 expression blendshapes, to the MediaPipe landmarks as participants performed different facial expressions.

We trained a deep neural network comprising per-grid spatial encoders followed by a dilated temporal convolutional network (TCN) to predict blendshape parameters from HD-sEMG signals at 100 Hz. Once trained, the network can predict expression blendshapes solely from HD-sEMG recordings. The output can be rendered using standard real-time blendshape animation methods. We demonstrate the methods using recordings from 25 participants, and direct expression transfer to a variety of human faces and non-human characters.

\end{abstract}
\section{Introduction}

Faces are central to human communication. Much of what we convey in conversation, from emotion and emphasis to attention and doubt, is carried by small movements of the face. As social interaction moves into virtual and augmented reality, capturing these movements faithfully has become an important problem in computer graphics.

Unfortunately, traditional methods for face capture are optical, requiring cameras and lighting to be mounted on a cumbersome rig held away from the face, that may interfere with other actors and the scene. The head-mounted displays that make such interaction immersive in XR also hide much of the upper face, and the inward-facing cameras that observe the eyes and mouth~\cite{Bai2024,Cheng2024,SamsungGalaxyXR}, see only oblique and incomplete views. Cameras pointed at the face also raise privacy concerns. Moreover, video sees only the result of muscle activity, not the activity itself, and muscle activations that barely deform the face are invisible to it~\cite{Gat2022}. We will refer to such optical devices, including traditional face capture devices and head mounted XR displays, as Head Mounted Devices in this paper (HMDs).

Here we take a different approach, and measure the cause of facial movement rather than its effect. We record high-density surface electromyography (HD-sEMG) with two textile grids of 32 electrodes each: one on the forehead, precisely where an HMD rests, and one on the cheek. From these signals alone, a neural network predicts the expression weights of a high-resolution parametric head model, GNM~\cite{Ploumpis2026}, at 100~Hz. Making this work required solving three problems: synchronizing EMG and video recorded by devices with independent clocks; encoding expressions from monocular video in a form that is independent of identity and pose; and learning the mapping from muscle activity to facial motion. We demonstrate the method on recordings of 25 participants.

\paragraph*{Our contributions:}
\begin{itemize}
  \item A capture setup and protocol for recording HD-sEMG from textile electrode grids together with video of the face.
  \item A novel synchronization method based on audio tones recorded by both devices, which models the drift between the two clocks and aligns the recordings to 0.15~ms RMS.
  \item A staged fit of a parametric head model to monocular landmarks, which separates head pose, eyelid closure and gaze, and the remaining expression, and yields expression weights that are independent of identity.
  \item A network with a spatial encoder for each grid and a dilated temporal convolutional network, which predicts expression weights from HD-sEMG alone. On held-out expressions, the median vertex error is 0.43~mm, about half that of a face held at the mean expression, and the eye region, which an HMD hides, is predicted as well as the rest of the face.
  \item An ablation showing that the forehead grid alone, which lies in the area of the HMD's face gasket, retains more than four fifths of the variance of facial motion explained with both grids.
  \item Demonstrations of expression transfer to other faces and to non-human characters, of decoding with the face covered, and of expressions and affect that are difficult to detect in video.
\end{itemize}

\section{Related Work}

\paragraph*{Face capture under occlusion.}
HMDs occlude much of the face. Li et al.~\cite{Li2015} combined strain gauges in the foam liner of an HMD with a camera on the mouth to drive a personalized blendshape model. Olszewski et al.~\cite{Olszewski2016} regressed the weights of a rig with 57 blendshapes from mouth and infrared eye cameras, and Wei et al.~\cite{Wei2019} animated person-specific photorealistic avatars from three headset cameras. Commercial headsets now include inward-facing cameras. The four cameras of the Meta Quest Pro drive universal codec avatars~\cite{Bai2024} and expression weights based on the Facial Action Coding System~\cite{MetaFaceTracking}; Apple Vision Pro tracks the face and eyes to animate a ``persona'' captured in advance~\cite{Cheng2024}; and Samsung's Galaxy XR has four eye-tracking cameras~\cite{SamsungGalaxyXR}. All of these rely on oblique, partial views from cameras held away from the face, which also raise privacy concerns.

\paragraph*{Facial muscles and surface EMG.}
Facial muscles insert into the skin and are interwoven, so they act more as a network than as independent actuators~\cite{Schumann2021,Mueller2022}. Surface EMG (sEMG) has traditionally been recorded over a few selected muscles, but recent work records many channels: 90 electrodes over the forehead and cheek of healthy subjects and patients with Bell's palsy~\cite{Cui2021}, bilateral recordings of mimic movements~\cite{Mueller2022} and of the six basic emotional expressions~\cite{GuntinasLichius2023}, and a 16-channel dry, skin-conformal array~\cite{Man2025}. Each electrode records a mixture of signals from overlapping muscles, and locating their sources is an ill-posed inverse problem~\cite{vandenDoel2008,vandenDoel2011}; the choice of signal processing also matters~\cite{Rutkowska2024}. Compared with action units detected in video, sEMG agrees on the timing of events but not consistently on their source, and it detects activations, such as clenching the teeth, that barely deform the face~\cite{Gat2022}.

\paragraph*{Decoding facial EMG.}
Most applications classify discrete states: facial gestures for hands-free control, from electrodes on a VR headset frame~\cite{Kim2023}; silent speech, from four 64-channel arrays~\cite{Chen2023}; or emotions, typically from two muscles~\cite{Rutkowska2024}. Continuous prediction of facial shape is much rarer. Eskes et al.~\cite{Eskes2017} predicted ten markers on the lips with a mean error of 2.76~mm. Man et al.~\cite{Man2025} predicted the 52 ARKit expression blendshapes (a mesh with 1,220 vertices~\cite{AppleARFaceGeometry}) for one participant, but performed poorly on a new one. EIFER~\cite{Buchner2025} restores faces occluded by electrodes using the FLAME model~\cite{Li2017} (5,023 vertices, 100 expression blendshapes), and maps between FLAME expression parameters and muscle activity; it observes that EMG underestimates mouth opening, which requires little muscle activity.

Brain signals have also driven avatars: Xiong et al.~\cite{Xiong2026} decoded scalp EEG into dense 3D faces. However, EEG is blurred by the skull and contaminated by facial muscle activity, whereas sEMG records directly the muscles that move the face.

Our work builds on these insights. We record 64 channels with two fabric grids of 32 electrodes, one on the forehead, which an HMD typically occludes; each grid is applied in a single step, which is much easier than placing individual electrodes. We fit the GNM head model~\cite{Ploumpis2026}, whose template mesh has 17,821 vertices, including the eyeballs, teeth, and tongue, and 383 expression blendshapes.

\begin{figure*}[t]
  \centering
  \setlength{\setupfigh}{0.345\textwidth}
  \begin{minipage}[b]{0.633\textwidth}
    \centering
    \includegraphics[height=\setupfigh]{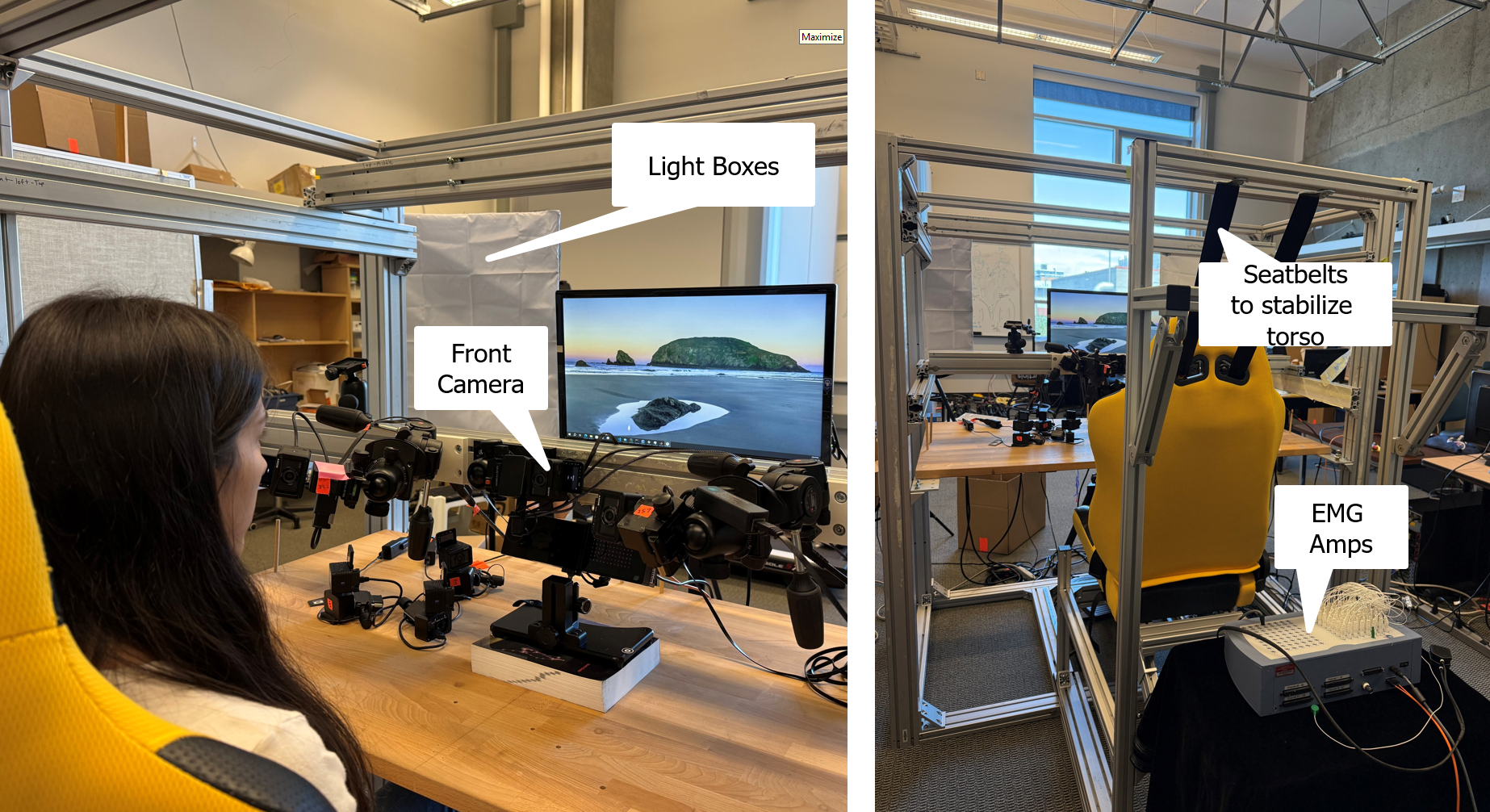}\\[2pt]
    {\small (a)}
  \end{minipage}\hfill
  \begin{minipage}[b]{0.348\textwidth}
    \centering
    \includegraphics[height=\setupfigh]{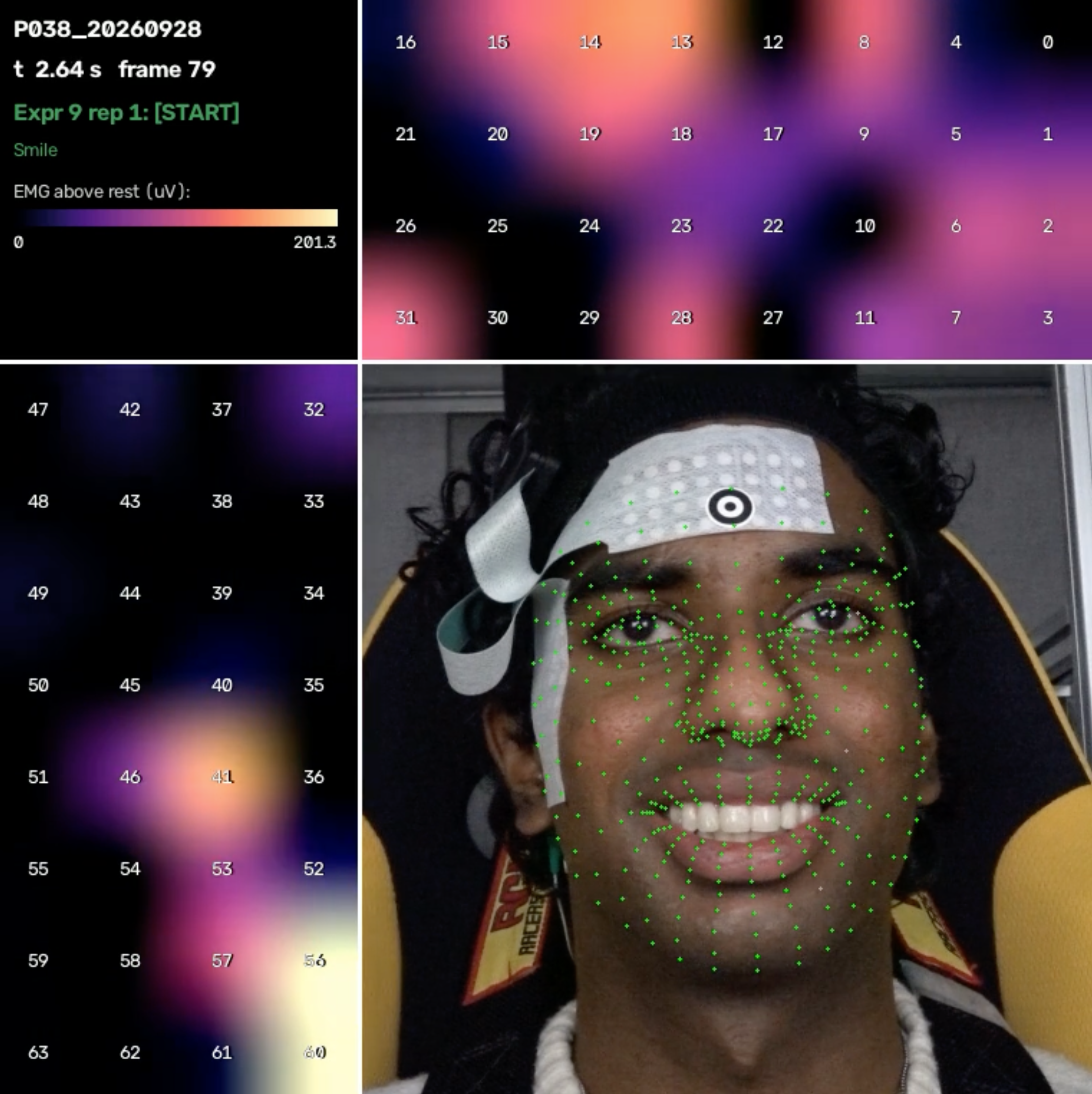}\\[2pt]
    {\small (b)}
  \end{minipage}
  \caption{(a) Capture setup. Left: the participant sits facing a monitor that displays instructions for each task, while a front camera records the face, which is lit by light boxes. Right: seatbelts attached to the chair stabilize the torso, and the EMG amplifiers are placed behind the chair. (b) Electrode grids placed on a participant's face (bottom right), with a visualization of the raw EMG signals recorded by the forehead grid (top) and the cheek grid (left). The numbers show the channel at each electrode.}
  \label{fig:setup}
\end{figure*}

\section{Methods}

\subsection{Data Acquisition}
\label{sec:acquisition}

\begin{figure*}[t]
  \centering
  \setlength{\setupfigh}{0.46\textwidth}
  \begin{minipage}[b]{0.49\textwidth}
    \centering
    \includegraphics[height=\setupfigh]{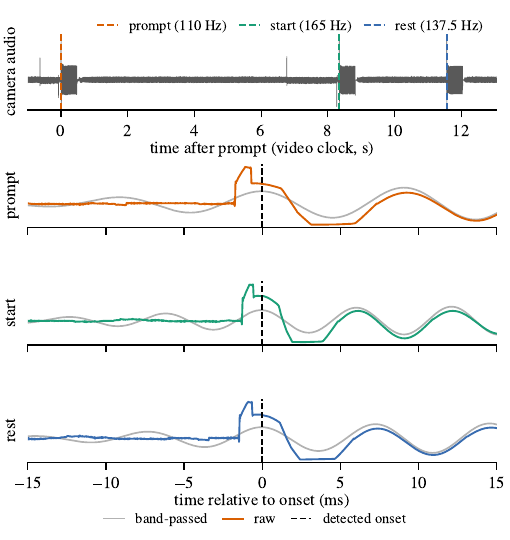}\\[2pt]
    {\small (a)}
  \end{minipage}\hfill
  \begin{minipage}[b]{0.49\textwidth}
    \centering
    \includegraphics[height=\setupfigh]{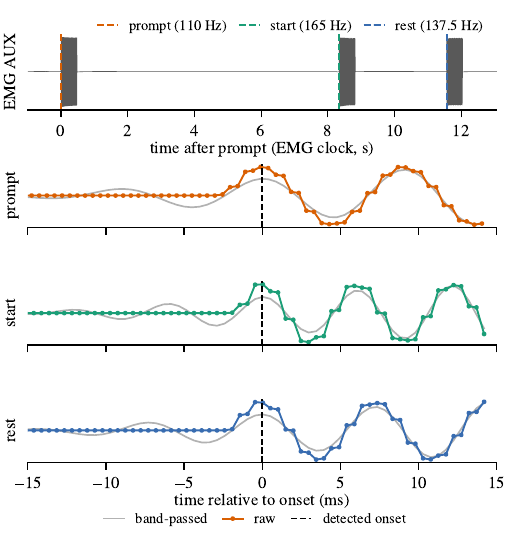}\\[2pt]
    {\small (b)}
  \end{minipage}\\[6pt]
  \includegraphics[width=0.85\textwidth]{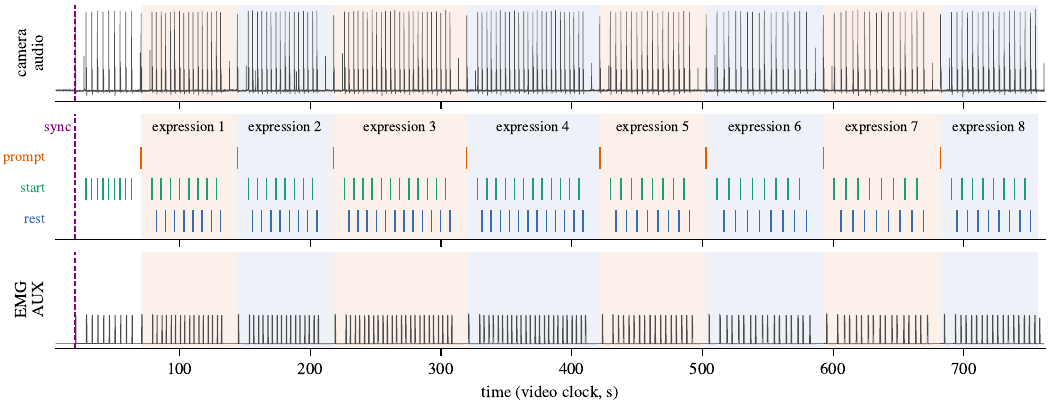}\\[2pt]
  {\small (c)}
  \caption{Synchronization and segmentation. (a) Tones detected in the audio track of the video. Top: the three tones that mark one trial. Below: each onset at millisecond scale, showing the raw audio sampled at 44.1 KHz, a band-passed copy (gray), and the detected onset (dashed). Note that the first cycle has some artifacts, likely due to the camera's gain control hardware, but our method is immune to such artifacts. (b) The same tones recorded on an auxiliary channel of the EMG amplifier, sampled at 2048~Hz. (c) One recording of eight expressions, preceded by saccade trials marked by start tones only: tone envelopes in the camera audio (top) and in the EMG channel (bottom), and the timeline of events recovered from them (middle). Note that the detected onset marks the same phase of the first cycle in both recordings, so any fixed bias cancels in the offset between the two clocks.}
  \label{fig:sync}
\end{figure*}

We developed a capture facility to record facial muscle activity and the resulting facial movements simultaneously, while the participant's torso movement is restrained to keep the head in the camera's field of view. Figure~\ref{fig:setup}a shows the capture setup.

\paragraph*{Participants.}
We report results from 25 participants, ages 19-70. All participants gave written informed consent. The protocol was approved by the University of British Columbia's Behavioral Research Ethics Board, in advance. Each session took about 60 minutes.

\paragraph*{Surface EMG.}
We recorded facial muscle activity with two 32-channel textile HD-EMG electrode grids (TMSi), one on the forehead, which is typically occluded by an HMD, and one on the upper right cheek (Figure~\ref{fig:setup}b). To reduce skin-to-electrode impedance, the skin was first prepared with a mildly abrasive gel (Nuprep) and cleaned with an alcohol swab. The electrodes were then filled with conductive gel (ECI Electro-Gel), and a disposable ground electrode was placed over the right mastoid process. The grids were connected to an ANT Neuro amplifier (Refa Ext) by shielded leads. A strap around the forehead (Artinis HD-EMG Mini Cradle Strap) holds the leads in place, and the leads were taped to the participant for strain relief, to reduce movement artifacts. The signals were digitized at 2048~Hz.

\paragraph*{Video.}
The face was recorded by a front-facing camera (Sony DSC-RX0M2, 3840x2160 pixels at 29.97 fps), mounted on a rigid aluminum frame (item Industrietechnik). The participant sat in a chair with seat belts to stabilize the torso.

\paragraph*{Protocol.}
Instructions for each task were shown on the monitor, with auditory cues marking preparation, movement onset, and rest. The tasks included targeted facial movements (e.g., raising and furrowing the eyebrows, blinking, and smiling) and counting to five. Each movement was repeated up to eight times, as naturally and consistently as possible. To minimize fatigue, participants rested for about 5~s between repetitions and 15~s between blocks of tasks.

\subsection{Synchronization}
\label{sec:sync}

Accurate synchronization is a crucial challenge for multi-modal capture. We learn a frame-by-frame mapping from muscle activity to facial motion, so any misalignment between the two recordings blurs the relation that the network must learn. However, the EMG amplifier and the camera sample at different rates (2048~Hz and 29.97 frames per second), using independent clocks that start at unknown times. We therefore align the two recordings using the one signal that both can record: sound. More precisely, we use audio-frequency tone bursts that are recorded electrically (without microphones and air propagation delay) by both the EMG digitizer and video camera's audio input jack.

The experiment computer marks every event of the protocol with a 0.5~s sine burst, whose frequency identifies the event: a prompt tone before each block of repetitions of an expression, a start tone at each movement onset, and a rest tone at each return to rest. A separate synchronization tone begins the recording. The same audio signal is recorded by cable, both on the camera's audio input and on an auxiliary analog channel of the EMG amplifier, so an experiment of eight expressions yields about 160 tones in each recording (Figure~\ref{fig:sync}).

We locate each tone onset to a fraction of a sample. We band-pass the recording around the tone's frequency, find candidate bursts from its envelope, and correlate each with a short template of the tone; parabolic interpolation of the correlation peak and the phase of the analytic signal give sub-sample precision. Band-pass filtering delays and smears the onset, so we finally step back one cycle at a time on the unfiltered recording until the waveform no longer matches the tone. This finds the first cycle of the burst (Figure~\ref{fig:sync}a,b). Since one EMG sample lasts 0.49~ms, and one audio sample far less, the onsets are located well below a millisecond in both recordings.

\begin{figure*}[t]
  \centering
  \includegraphics[width=\textwidth]{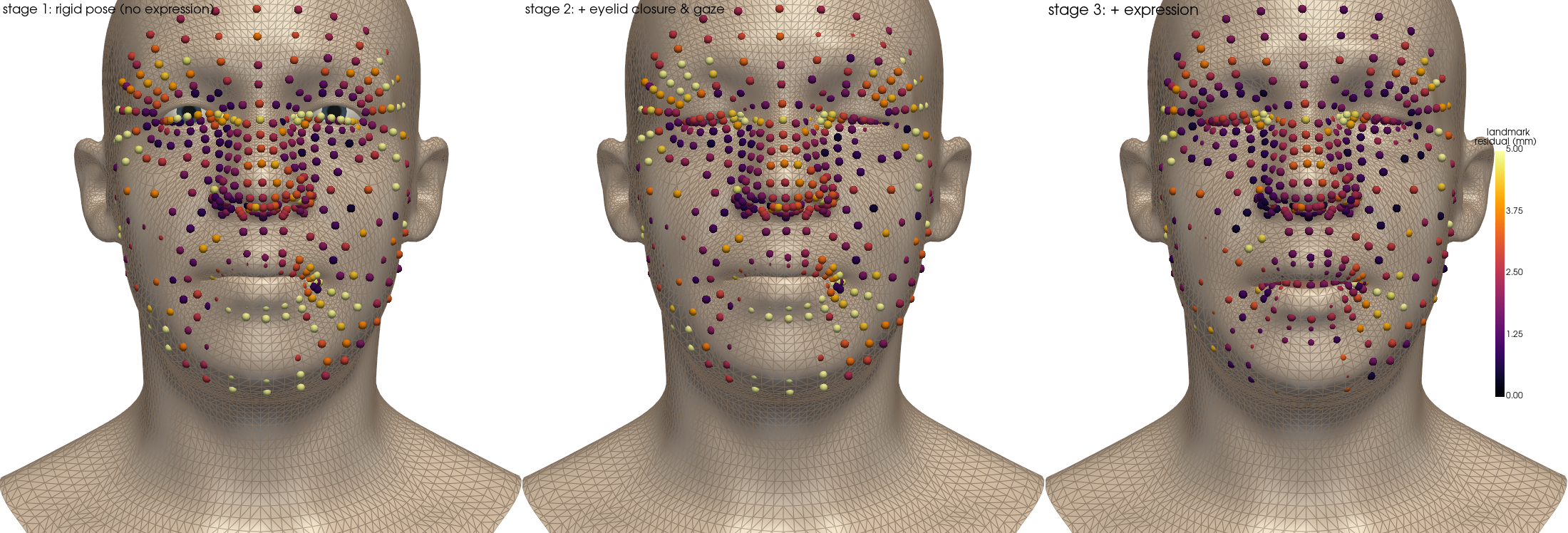}
  \caption{Staged fitting of GNM to MediaPipe landmarks, for a frame in which the participant blinks while speaking. Spheres show the landmarks, colored by their distance from the fitted mesh (0--5~mm). Left: stage 1 fits only the rigid pose of the head. Middle: stage 2 adds eyelid closure and gaze, which fits the closed eyes. Right: stage 3 adds the remaining expression blendshapes, which fits the open mouth and reduces the residuals over most of the face.}
  \label{fig:fitting}
\end{figure*}

The two clocks run at slightly different rates, so we map EMG time to video time with a linear function, $t_{\mathrm{video}} = (1+\delta)\,t_{\mathrm{EMG}} + t_0$, where $t_0$ is the offset between the clocks and $\delta$ their relative drift. The onsets of the synchronization tone give an initial estimate of the offset, which we use to match each tone in one recording to the same tone in the other. We then fit $t_0$ and $\delta$ by least squares to the onsets of all matched tones, after rejecting outliers. We segment the trials from the tone onsets and map the segments between the two clocks with this function (Figure~\ref{fig:sync}c). The EMG and the landmarks of each trial are then extracted on their own clocks. How accurate is this alignment? In a representative 12-minute recording, 161 tones were matched between the two devices, and the fitted drift is $\delta = 3.1$~ppm. This is small, but not negligible: a constant offset alone would misalign the recordings by up to 1.8~ms by the end of the recording. With the drift included, the matched onsets agree to 0.15~ms RMS (0.49~ms at most). Note that this measures the alignment of the two recordings of the tones; it does not include latencies inside the camera between video and audio streams, that we assume are negligible.

\subsection{Facial Expression Encoding}
\label{sec:features}

\begin{figure*}[t]
  \centering
  \includegraphics[width=\textwidth]{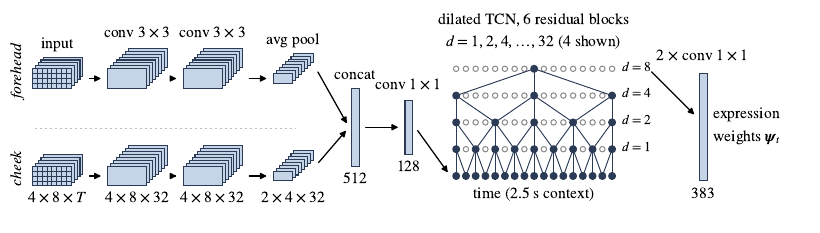}
  \caption{Network architecture. Each grid is a $4\times8$ image of EMG envelopes at every frame, and has its own spatial encoder of two $3\times3$ convolutions followed by average pooling. The pooled features of both grids are concatenated and projected to 128 features per frame. A dilated temporal convolutional network (TCN) then combines 2.5~s of context centered on each frame, and two $1\times1$ convolutions map the result to the 383 GNM expression weights. Note that the dilations double with depth, so six layers of kernel size 3 span the whole context.}
  \label{fig:architecture}
\end{figure*}

We encode facial expressions in two steps. We first extract facial landmarks from each video frame, and then encode the 3D geometry of the face using the GNM~\cite{Ploumpis2026} parametric head model.

In each video frame, we estimated 478 3D facial landmarks, including 10 on the irises, using MediaPipe's Face Landmarker~\cite{MediaPipeFaceLandmarker}. Since the landmarks are estimated from a single camera, their depth is much less accurate than their position in the image plane. These landmarks are therefore a sparse and noisy estimate of the shape of the face. More importantly, they are not a good target for learning, since their positions confound three distinct sources of variation: the participant's identity, the pose of the head, and the facial expression. Only the last is caused by the facial muscles. We therefore encode facial expression in the space of a parametric head model, GNM. GNM deforms a template mesh with 17,821 vertices by linear combinations of 253 identity components and 383 expression blendshapes, and poses it with a skeleton of four joints. We fit GNM to the MediaPipe landmarks in three stages (Figure~\ref{fig:fitting}).

We use three stages because the face is highly mobile, and a single fit to all landmarks would mix the effects of skeletal pose, gaze and blinks, and expression. To relate the landmarks to the model, we placed all 478 MediaPipe landmarks on the GNM template mesh by hand, following the numbered landmark map in the MediaPipe documentation~\cite{MediaPipeFaceLandmarker}. Before the stages, we compute a neutral face once per session, as the median of the landmarks over the rest phases. The first stage then finds the rotation and translation of the head in each frame by a Procrustes fit to 16 relatively stable landmarks, over bones of the upper face, excluding the irises, the eyelids, and the landmarks between and above the brows, which rise with them. The second stage reads the closure of each eye from its aperture landmarks, relative to its neutral opening, and smooths it over the whole trial without blurring the blinks; a closure curve calibrated for each session maps it to the 200 eye-region blendshapes of GNM. The gaze of each eye is fitted to the iris landmarks, which are down-weighted when the lid covers the iris, making gaze estimates robust to blink artifacts. Since MediaPipe shifts the whole mesh slightly when the eyes close, we then subtract a per-session estimate of this shift from the landmarks outside the eye sockets. The third stage fits the remaining expression to all landmarks except the eyelids. It uses the other 183 blendshapes, and the eye-region blendshapes restricted to the null space of their displacements at the eyelid margins, which we compute by singular value decomposition (116 of 200 dimensions). The brows and the forehead can then move, while the eyelids stay where the second stage put them. Since the depth of the landmarks is unreliable, this stage also holds the forehead and brows at the depth of the neutral face.
Since the identity parameters are shared by all frames of a participant, the fitted expression weights $\boldsymbol{\psi}_t \in \mathbb{R}^{383}$ at frame $t$ describe the expression independently of identity and head pose. These weights are resampled to 100~Hz by linear interpolation (Appendix~\ref{sec:emg-processing}), synchronized with the EMG (Section~\ref{sec:sync}), and serve as the targets for learning. Encoding expressions as blendshape weights has two further advantages. The predicted weights can be rendered by standard real-time blendshape animation, and they can be applied directly to other identities and characters (Section~\ref{sec:transfer}).

\subsection{Predicting Expressions from HD-sEMG}
\label{sec:learning}

The raw sEMG cannot be used directly for learning, since it is dominated by large offsets and slow drift rather than by muscle activity. We therefore apply several signal processing steps, which convert each of the 64 channels into an envelope of muscle activation sampled at 100~Hz (see Appendix~\ref{sec:emg-processing} in the Supplementary Materials).

We train a separate network for each participant to map the EMG envelopes (Appendix~\ref{sec:emg-processing}) to the fitted expression weights $\boldsymbol{\psi}_t$. A network per participant is natural here, since the grids are placed by hand and the same electrode lies over slightly different muscles in each session.

The network has two parts (Figure~\ref{fig:architecture}). First, a spatial encoder treats each grid as a $4\times8$ image of electrode activity and applies two $3\times3$ convolutions to it. The two grids lie over different muscles, so each has its own encoder, and its output is pooled to $2\times4$ to retain which part of the grid is active. Second, a temporal convolutional network (TCN)~\cite{Bai2018} with six residual blocks, whose dilations double from 1 to 32, combines 2.5~s of context centered on each frame. The network has about 0.75 million parameters. Each input window is normalized per channel~\cite{Ulyanov2016}, which removes differences in electrode gain between channels and sessions. The targets are shifted by a single time lag per participant, estimated by cross-correlating the EMG with the motion of the fitted weights (median 70~ms). We minimize a Huber loss~\cite{Huber1964} on the weights and on their frame-to-frame differences with AdamW~\cite{Loshchilov2019}, for up to 150 epochs, and keep the network that best predicts a validation set. Training takes between 4 and 60 minutes per participant on a single GPU.

For testing, one repetition of every expression in each recording block, chosen at random, is held out (16 of about 145 trials per participant). This list of test trials is fixed once for all experiments, and 15\% of the remaining trials are used for validation.

\subsection{Correcting Eyelid Intersections}
\label{sec:eyelids}

\begin{figure}[ht]
  \centering
  \includegraphics[width=\linewidth]{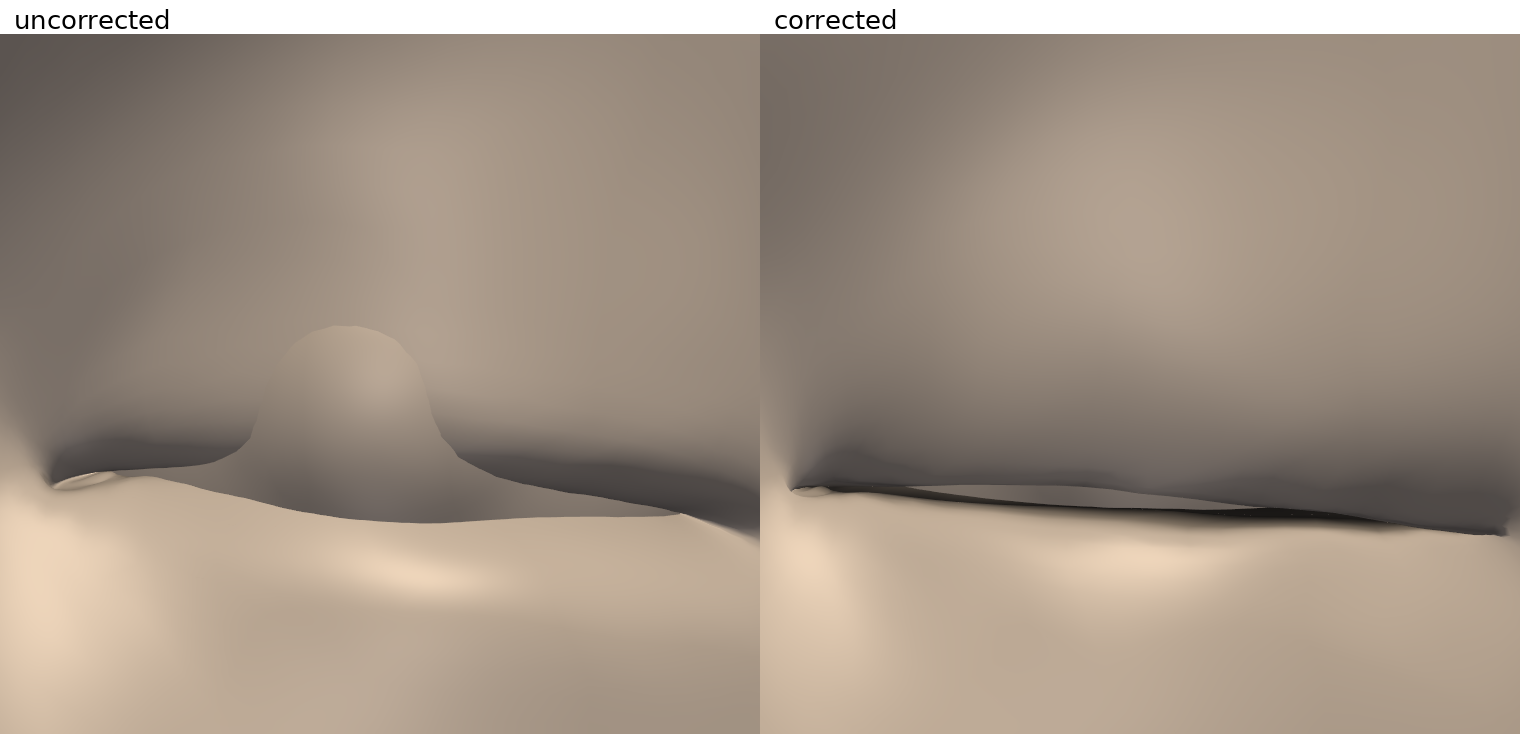}
  \caption{A closed eyelid before (left) and after (right) the correction, on the same frame. Left: the eyeball emerges through the lid, which penetrates it by up to 3.4~mm. Right: after Equation~\ref{eq:eyelid}, nothing penetrates, and the margin of the lid reads as a closed crease.}
  \label{fig:eyelids}
\end{figure}

The eye stage of our fit (Section~\ref{sec:features}) is deliberately unconstrained, since a constrained fit damps exactly the large weights that a blink needs, and leaves the lids open. However, nothing in the blendshape model represents the eyeball, so weights that reproduce the measured eye aperture are equally free to drive the lid through the globe; at full closure, the lid penetrated it by up to 3.4~mm (Figure~\ref{fig:eyelids}, left). Following Neog et al.~\cite{Neog2016}, we constrain the lid against the eye. We fit a sphere to the vertices of the sclera (radius 14.5~mm), and define a target shell of radius $R$, one skin thickness $\tau$ outside it. An eyelid vertex at distance $\rho$ from the center $\mathbf{c}$, in direction $\mathbf{u}$, is moved to
\begin{equation}
  \mathbf{x}' = \mathbf{c} + f(\rho)\,\mathbf{u}, \qquad
  f(\rho) = R + \sigma \log\!\left(1 + e^{(\rho - R)/\sigma}\right).
  \label{eq:eyelid}
\end{equation}
Note that $f$ is strictly increasing, $f(\rho) > R$ everywhere, and $f(\rho) \to \rho$ for $\rho \gg R$: nothing penetrates the shell, geometry away from the eye is unchanged, and distinct radii remain distinct.
We use $R = 1.5$~mm and $\sigma = 2$~mm, and correct a fixed set of vertices for the whole sequence, to avoid temporal flicker. Shading normals are transformed consistently with the map.

\section{Results}

We recorded HD-sEMG and video from 25 participants as they performed 8 facial expressions, mostly involving the upper face (e.g., brow raise/blink), plus smile and speech (reciting numbers 1 to 5). In this section we first evaluate how well facial expressions can be predicted from HD-sEMG alone. We then demonstrate applications that are difficult or impossible with optical face capture. Unless stated otherwise, all results are computed on the held-out trials of each participant (Section~\ref{sec:learning}). Please see the accompanying video for animated results.

\subsection{Performance and Validation}
\label{sec:performance}

How well do predictions from HD-sEMG match the expressions tracked from video? We use two measures, computed on the held-out trials (Section~\ref{sec:learning}). The first is the Pearson correlation $r$ between the predicted and fitted weights of each GNM expression blendshape, averaged over the blendshapes that vary in the test data. However, the blendshapes are far from equal: a unit weight moves the face by amounts that differ by a factor of about 500 between blendshapes.
Our second measure is computed on the mesh itself. We reconstruct the GNM mesh from the predicted and from the fitted weights, and report the mean distance between corresponding vertices of the two meshes and the fraction $R^2$ of the variance of the fitted facial motion that the prediction explains. Unlike $r$, this weights each blendshape by how much it actually moves the face. We compute these and all other measures on the mesh over the visible face only: the 5,101 vertices of exterior skin that expression can move. The expression blendshapes also move the teeth, gums, and tongue, the inside of the mouth, and the eye sockets. No camera sees these parts and no landmark constrains them, so we exclude them, together with the eyeballs, which only rotate with gaze. Note that the fitted mesh is itself an estimate, so these measures are relative to the video-based fit and not to the true shape of the face.

\begin{figure}[ht]
  \centering
  \includegraphics[width=\linewidth]{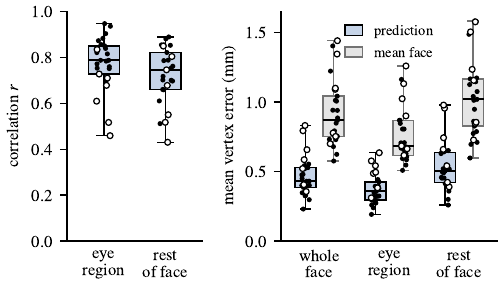}
  \caption{Prediction performance across participants, on the held-out cued expressions. Each box shows the distribution over the 25 participants, with open circles marking the seven recordings of lower signal quality. Left: correlation between predicted and fitted blendshape weights, for the eye-region blendshapes of GNM (in GNM, the eye region includes the forehead) and for the rest of the face (higher is better). Right: mean vertex error of the reconstructed mesh, over the whole face, in the eye region, and elsewhere, next to the error of holding each participant's mean face (gray; lower is better). Note that the error is about half that of the mean face in every region.}
  \label{fig:perf-summary}
\end{figure}

Figure~\ref{fig:perf-summary} summarizes performance across participants, on the cued expressions. The median correlation is $r = 0.76$ (range 0.48--0.92 over the 25 participants), and the predictions explain a median of 64\% of the variance of facial motion (range 20--91\%). The median vertex error is 0.43~mm (range 0.23--0.83~mm), about half the 0.87~mm error of a face that holds each participant's mean expression throughout. The eye region is predicted as well as the rest of the face. Its blendshapes correlate with a median $r$ of 0.79 (range 0.46--0.95), against 0.75 elsewhere, and its error is again about half that of the mean face. The network outperforms a linear model (ridge regression on the envelopes at five time lags up to 400~ms) for 23 of the 25 participants, with a mean $r$ of 0.67 against 0.55 over all test trials.

Why do some participants score much lower than others? Most of the low scores come from seven participants whose recordings had an elevated EMG level at rest, or little difference in level between task and rest (mean $r = 0.65$, against $0.79$ for the other 18). In these recordings the predictions explain 54\% of the variance, against 66\% for the other participants, and miss most of the blinks (Appendix~\ref{sec:interpreting}). This is consistent with poor electrode contact, and suggests that checking signal quality at capture time is as important as the choice of network.

\begin{figure}[ht]
  \centering
  \includegraphics[width=\linewidth]{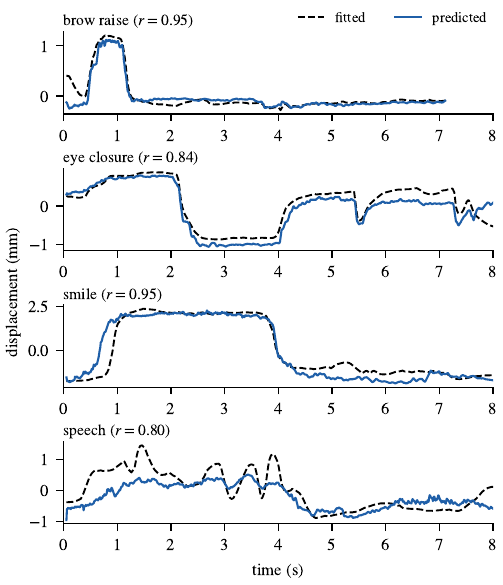}
  \caption{Predicted (solid) and fitted (dashed) facial motion during four held-out trials of one participant, whose accuracy on these expressions is close to the median over all participants. Each trace is the displacement of the face along the main mode of motion of that trial, and $r$ is computed over the whole trial. For most participants brow raises, eye closure, and blinks are tracked closely, including their timing, while the lip and jaw motion of speech is underestimated, as B\"uchner et al.~\cite{Buchner2025} also observed.}
  \label{fig:traces}
\end{figure}

Figure~\ref{fig:traces} shows the predicted and fitted motion over time, for four held-out trials of one participant. To compare faces rather than individual weights, we project both onto the main mode of motion of each trial, i.e., the displacement pattern that accounts for most of the fitted motion. The onsets and peaks of brow raises, eye closure, and blinks agree closely with the video. Speech is the weakest case: its lip and jaw motion is underestimated. This is consistent with the observation of B\"uchner et al.~\cite{Buchner2025} that expressions synthesized from muscle activity underestimate expressions that open the mouth, since holding the mouth open utilizes gravity and requires little muscle activity.

\begin{figure}[ht]
  \centering
  \includegraphics[width=\linewidth]{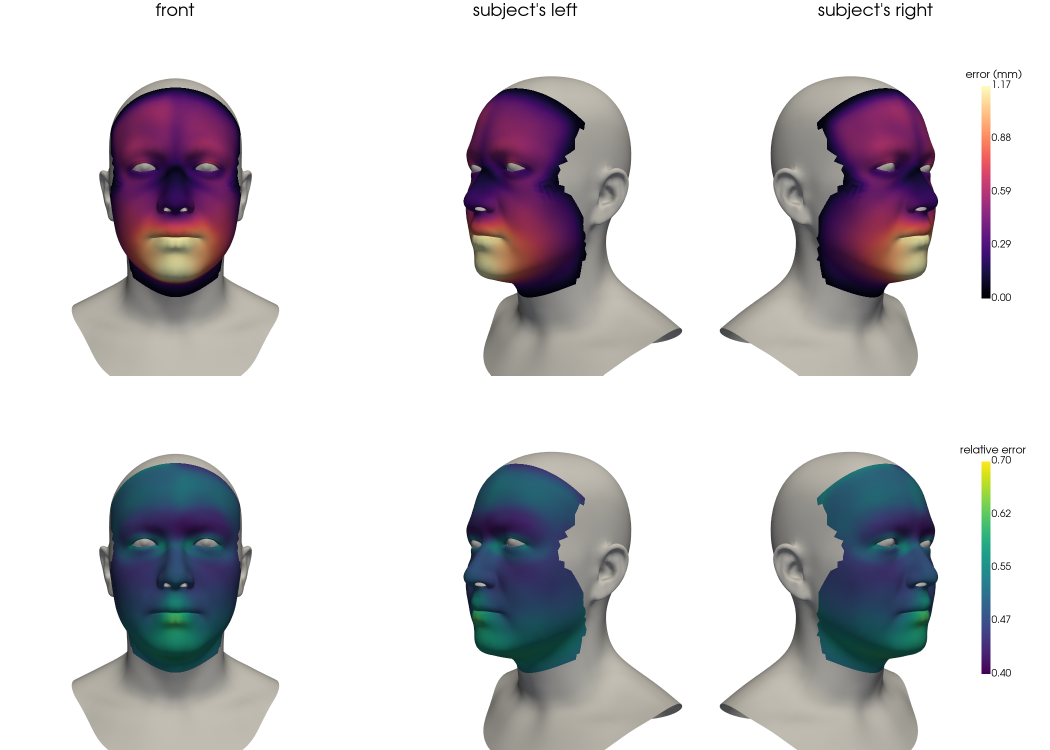}
  \caption{Where the errors are. Top: mean vertex error (mm) on the held-out cued expressions, averaged over the 25 participants and shown on the GNM template from the front and from both sides. Bottom: the same error relative to the error of holding each participant's mean face (lower is better). Vertices that expression cannot move are gray. Note that the relative error is lowest at the brows and the cheeks and highest at the chin, and that both maps are nearly symmetric.}
  \label{fig:error-map}
\end{figure}

Where on the face are the errors largest? Figure~\ref{fig:error-map} shows the mean vertex error on the GNM template. In absolute terms, the error is largest at the lips and the chin (up to 1.25~mm), which also move the most, and smallest on the nose and at the sides of the face. Relative to the error of holding the mean face, the eye region and the rest of the face are predicted equally well (a median ratio of 0.50 in both). Interestingly, the error is the same on both sides of the face (0.47~mm), although the cheek grid lies on one side only. This is perhaps because the cued expressions move both sides of the face together.

These numbers must be interpreted with caution. Appendix~\ref{sec:interpreting} in the Supplementary Materials examines the subtleties, including the accuracy of the eyelids while they move.

\subsection{Decoding with the Face Covered}
\label{sec:covered}

\begin{figure}[ht]
  \centering
  \includegraphics[width=\linewidth]{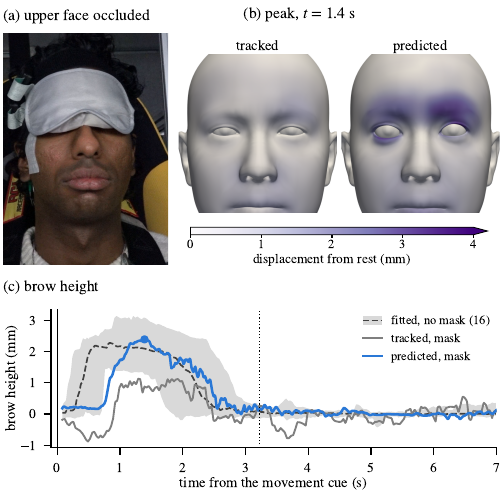}
  \caption{Decoding facial expressions with the face covered by a sleep mask covering the forehead. (a) Upper face occluded. (b) GNM faces at the peak of one brow raise under the mask, fitted to the landmarks tracked by MediaPipe and predicted from HD-sEMG, colored by displacement from rest. (c) Brow height over the same trial; the dashed line and gray band show the median and range of the participant's 16 brow raises without the mask, and the dotted line the rest cue. Note that the tracked brows rise only half as far, while the predicted brows rise as far as without the mask.}
  \label{fig:covered}
\end{figure}

Optical face capture fails when the face is covered, for example by an HMD or a mask, but HD-sEMG needs no view of the face. We recorded one more participant in three blocks; in the third, the participant wore a sleep mask over the eyes and the forehead, and thus over the forehead grid (Figure~\ref{fig:covered}a). We trained the network on the two uncovered blocks only. On held-out repetitions of those blocks it explains 69\% of the variance of facial motion ($r = 0.82$). Under the mask, MediaPipe still reports a face in 99\% of the frames, but it draws the eyes and the brows onto the mask. Its eyelids never blink, and its brows rise by only 1.0~mm during brow raises, against 2.3~mm without the mask. The EMG, in contrast, is unaffected. The predicted brows rise by 1.9--2.7~mm in seven of the eight brow raises, within the range without the mask (1.4--3.1~mm; Figure~\ref{fig:covered}b,c), and the predicted eyes close in all 12 cued eye closures. We cannot measure the accuracy of these predictions, since the mask hides the face we would compare them with.

\subsection{Decoding from the Forehead Alone}
\label{sec:forehead}

The forehead grid lies where the face gasket of an HMD rests. How much of the accuracy above does it provide by itself? To find out, we trained the network for every participant without the cheek grid: its spatial encoder is removed, and the time lag is estimated from the forehead grid alone. Everything else, including the held-out trials, is unchanged.

\begin{figure}[ht]
  \centering
  \includegraphics[width=\linewidth]{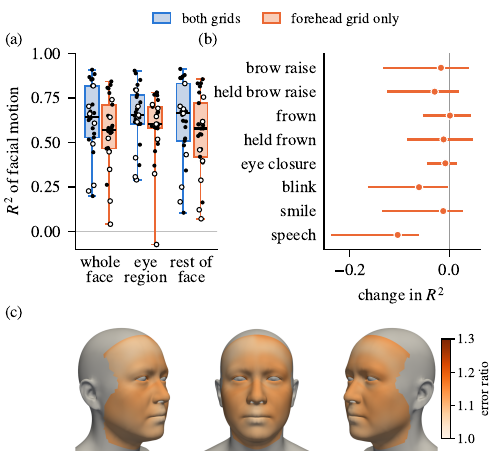}
  \caption{Decoding from the forehead grid alone. (a) Fraction $R^2$ of the variance of facial motion explained for each participant, with both grids and with the forehead grid alone, over the whole face, the eye region (which in GNM includes the forehead), and the rest of the face. Open circles mark the seven recordings of lower signal quality. (b) Change in $R^2$ over the frames of each cued expression when the cheek grid is removed (median and interquartile range). (c) Mean vertex error with the forehead grid alone, relative to that with both grids, from the subject's right, the front, and the subject's left. Note that the error grows by a similar fraction over the whole face.}
  \label{fig:forehead}
\end{figure}

With the forehead grid alone, the median correlation drops from 0.76 to 0.72, and the predictions explain a median of 57\% of the variance of facial motion, against 64\% with both grids (Figure~\ref{fig:forehead}a); 17 of the 25 participants are predicted less well. The median vertex error rises from 0.43 to 0.50~mm, still well below the 0.87~mm error of the mean face, and the network still outperforms a linear model on the same electrodes for every participant. Retraining both networks with another random seed gives a larger loss (a median of 66\% against 53\%).

Where is the accuracy lost? The rest of the face loses more than the eye region: its $R^2$ drops from 0.66 to 0.58, against 0.65 to 0.61 in the eye region. However, the error grows by a similar fraction at every vertex (a median of 10\%), and equally on both sides of the face, although the cheek grid lies on the right side only (Figure~\ref{fig:forehead}c). Thus the cheek grid does not simply report on the muscles beneath it. Rather, both grids appear to tell the network which expression is under way and how it unfolds, and the network reproduces the motion of the whole face from this. This would also explain why the forehead grid, which records no muscle of the lips, still accounts for about half of the variance of motion outside the eye region. Speech loses the most, with a median $R^2$ of 0.42 against 0.62 (Figure~\ref{fig:forehead}b).

The forehead grid alone therefore retains more than four fifths of the variance explained with both grids, which makes it a promising sensor for an HMD. However, the cheek grid improves every part of the face, and speech in particular.

\subsection{Decoding Affect and Invisible Expressions}
\label{sec:clench}

Facial muscles reflect a person's emotional state, even when the resulting expression is too subtle to see, or is deliberately suppressed. This is why facial EMG is widely used in psychophysiology to detect emotions from subtle facial expressions~\cite{Rutkowska2024}. An avatar driven only by visible deformation of the face cannot convey such states. Since HD-sEMG measures muscle activity directly, it can.

\begin{figure}[ht]
  \centering
  \includegraphics[width=\linewidth]{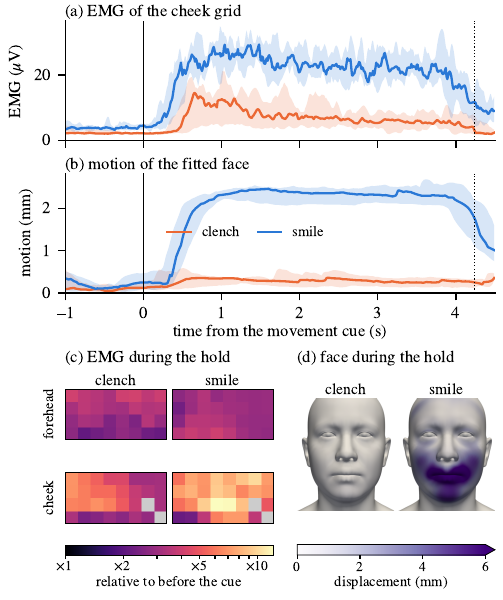}
  \caption{Clenching the teeth, an expression that barely moves the face. One participant clenched the teeth (orange) and smiled (blue) in seven repetitions each. (a) EMG of the cheek grid: envelope of the differences between neighboring electrodes, median over the pairs of electrodes. Lines show the median over repetitions and bands their range; the movement cue is at 0~s and the rest cue at the dotted line. Panels (b)--(d) are relative to the second before the cue. (b) Motion of the fitted GNM face: mean distance between corresponding vertices. (c) EMG of each electrode during the hold, laid out as in Figure~\ref{fig:emg}f (Appendix~\ref{sec:emg-processing}); the two faulty electrodes (gray) are excluded. (d) Displacement of the fitted face during the hold. Note that the clench produces strong EMG, in a pattern of its own, while the face hardly moves.}
  \label{fig:clench}
\end{figure}

Here we give one example: clenching the teeth, a familiar sign of effort, stress, or suppressed anger. Clenching is nearly isometric. Once the teeth meet, the jaw cannot close any further, so the jaw muscles tense without moving it (you can feel this with a fingertip on the side of your jaw). Gat et al.\ note that video analysis is insensitive to such activations~\cite{Gat2022}. We recorded one further participant, not among the 25 above, whose second recording block included clenching the teeth eight times in succession, for 4.2~s at each cue. We measure every change between the second before the cue and the hold, from 0.5~s after the cue until the rest cue, and compare the clenches with the same participant's smiles (Figure~\ref{fig:clench}). The first repetition shows no sign of a clench in either the EMG or the tracked face, and we analyze the other seven.

The face barely moves (Figure~\ref{fig:clench}b,d). During the clench, the vertices of the fitted GNM face move by 0.27~mm on average (median over repetitions; range 0.18--0.36~mm). This is little more than in the seconds of rest after a single blink (0.21~mm), an eighth of the motion of a smile (2.27~mm), and less than the median error of our predictions (0.43~mm, Section~\ref{sec:performance}). The largest displacement, 0.9~mm, is at the chin, which rises by about 1~mm in the tracked landmarks as the teeth meet. An avatar animated from the fitted weights, or from weights predicted to match them, would therefore show almost nothing.

In contrast, the EMG responds strongly. The envelope of the cheek grid rises to four times its level before the cue (median over its electrodes and over repetitions; Figure~\ref{fig:clench}c), against seven times during a smile. This activity is local. It persists, at three times the level before the cue, in the differences between neighboring electrodes, which cancel any signal common to both (Figure~\ref{fig:clench}a). It is also concentrated at one edge of the grid, in a pattern distinct from that of the smile. The forehead grid also records the clench, but almost uniformly: its envelope rises by a factor of 3.3, and the differences between its electrodes by only 1.3. The cheek grid behaves in the same way during a held brow raise (3.4 and 1.0), as expected of activity conducted from muscles that do not lie under the grid. On the cheek grid, the activity peaks as the teeth meet, and falls by about half from the first to the last second of the hold. Over the same interval the fitted face changes by only 0.19~mm, less than at rest after a blink (0.25~mm).

Can the clench be decoded? We trained a linear classifier (logistic regression) on the envelopes of the 62 working electrodes to separate the frames of each clench from rest and from the participant's other expressions, holding out one repetition of every expression at a time. It separates them almost perfectly: the area under the receiver operating characteristic (ROC) curve is 0.999, and at a threshold that flags 1\% of the other frames, it flags at least 96\% of the frames of every held-out clench. The forehead grid alone, which an HMD would cover, still reaches 0.99, although it misses much of the two weakest clenches. Note, however, that clenching is not entirely invisible: the rise of the chin is consistent enough for the same classifier to detect the clench from the tracked landmarks as well (0.999). The face thus shows that the teeth are closed, but little of the muscle activity that keeps them closed, which HD-sEMG records directly.

This evidence comes from a single participant, who clenched on instruction rather than under stress. Nevertheless, it shows that an expression can involve strong and repeatable muscle activity while moving the face by less than the error of our predictions.

\subsection{Expression Transfer}
\label{sec:transfer}

\begin{figure}[ht]
  \centering
  \includegraphics[width=\linewidth]{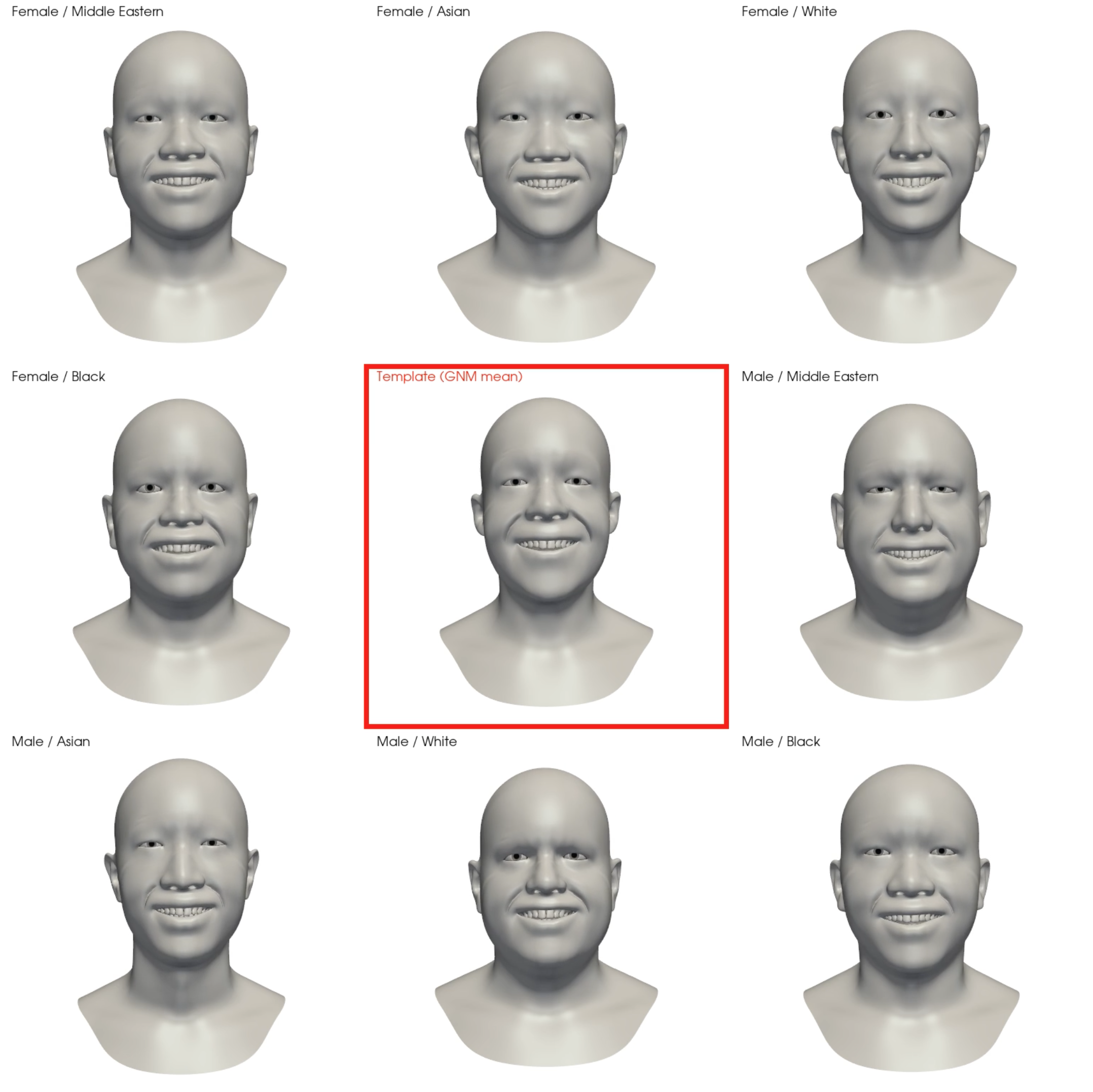}
  \caption{The same expression weights applied to nine GNM heads: the mean head (center, outlined) and eight other identities.}
  \label{fig:transfer-gnm}
\end{figure}

Since the network predicts GNM expression weights, which are independent of identity, the same expression can be applied directly to any GNM head. Figure~\ref{fig:transfer-gnm} shows two frames of one trial, in which the participant opens the eyes wide and then closes them, applied to the mean GNM head and to eight other GNM identities of both sexes.
Note that the eyes move consistently with the face, and that the forehead wrinkles (Section~\ref{sec:wrinkles}) appear on every head.

\begin{figure}[ht]
  \centering
  \includegraphics[width=\linewidth]{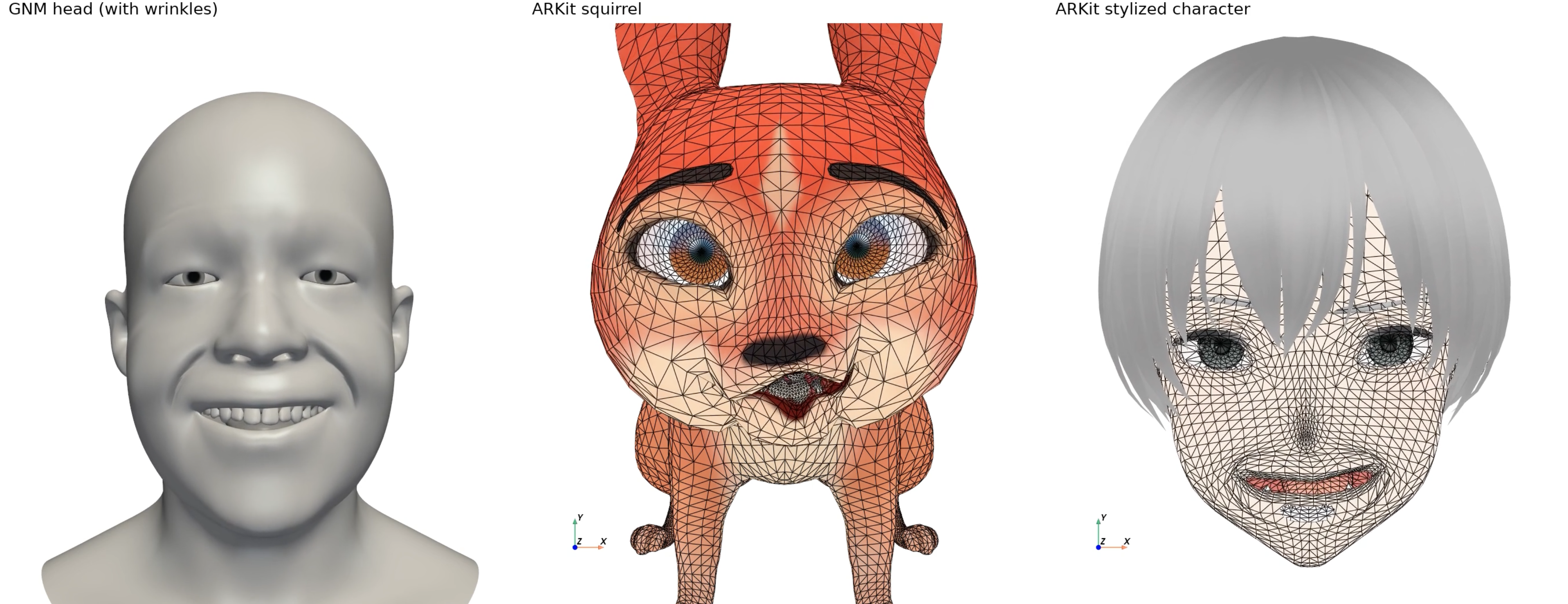}
  \caption{Expressions transferred to non-human characters. Expression weights from the GNM head (left) are transferred to multiple non human characters: a squirrel (middle) and a stylized character (right). Current limitations on registration for extremely different facial types result in artifacts for the squirrel character while smiling.}
  \label{fig:transfer-nonhuman}
\end{figure}

The predicted weights can also drive characters that are not human.
Figure~\ref{fig:transfer-nonhuman} shows characters animated from the same HD-sEMG recordings.

\subsection{Adding Secondary Effects}
\label{sec:wrinkles}

\begin{figure}[ht]
  \centering
  \includegraphics[width=\linewidth]{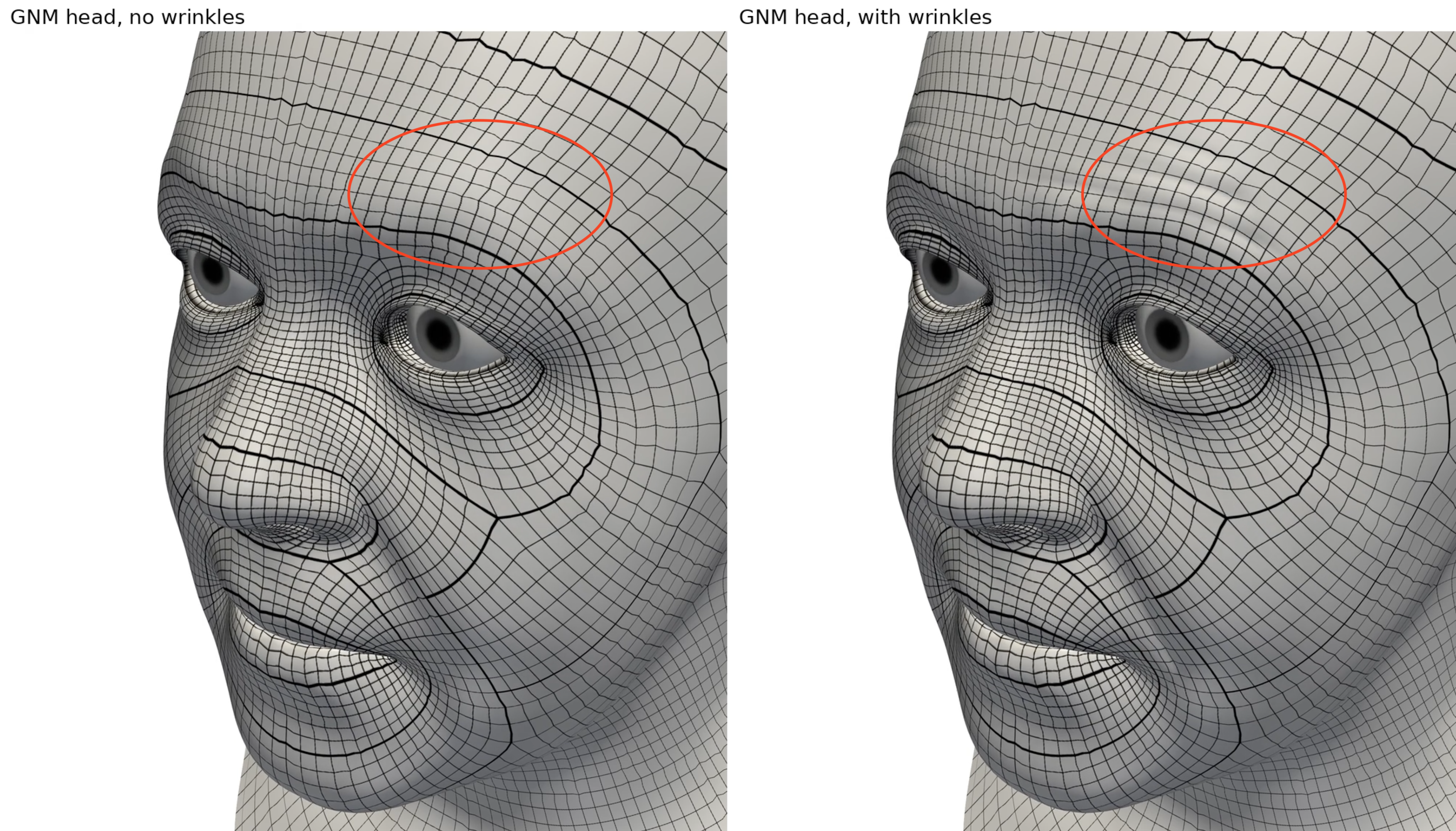}
  \caption{Without (left) and with (right) expression-driven wrinkles, on the same frame. Observe the highlighted region, as the subject raises their eyebrows furrows are naturally created along the forehead, through Equation~\ref{eq:wrinkles}.}
  \label{fig:wrinkles}
\end{figure}

Since we reconstruct a high-resolution skin mesh from sparse landmarks, we can apply secondary effects based on skin deformation, such as wrinkles. The GNM expression blendshapes are smooth, and do not reproduce the furrows that form on the forehead when the brows are raised. We therefore add large-scale expressive wrinkles following Bando et al.~\cite{Bando2002}, driven by the expression weights of each frame. Each wrinkle is a cubic B\'ezier curve in the texture domain of GNM; we use three horizontal furrows on the forehead and a pair of nasolabial folds. A vertex at distance $l_k$ from the center line of furrow $k$, measured on the rest surface, is displaced along its normal by
\begin{equation}
  \begin{gathered}
    \Delta = \sum_k S(l_k)\,\max(s_k, 0), \qquad s = 1 - \|\mathbf{F}\mathbf{p}\|, \\
    S(l) = d \left(\frac{l}{w} - 1\right) e^{-l/w}.
  \end{gathered}
  \label{eq:wrinkles}
\end{equation}
Here $S$ is Bando's profile, with depth $d$ and width $w$. Since $\int_0^\infty S(l)\,dl = 0$, the volume displaced into the furrow equals the volume raised beside it, so the skin is not compressed. The amplitude is set by the shrinkage $s$ of the skin across the furrow, where $\mathbf{F}$ is the deformation gradient of the surface relative to the participant's own rest pose, and $\mathbf{p}$ is the unit tangent perpendicular to the furrow. A furrow therefore deepens only when the skin is compressed across it, and not when it is deformed along it. This distinguishes our formulation from scalar measures of mesh tension~\cite{Raman2023}, which average over all edges at a vertex and so cancel under anisotropic deformation. We use $d = 1.3$--$1.7$~mm and $w = 4.5$--$5.0$~mm on the forehead, and $d = w = 4$~mm for the nasolabial folds, and scale $s$ by 15, since the physiological strain of a few percent alone produces a barely visible furrow (Figure~\ref{fig:wrinkles}).

\section{Conclusions}

We have shown that high-density surface EMG, recorded by two textile electrode grids on the forehead and the cheek, can drive a high-resolution parametric model of the face. Recording muscle activity and facial motion together, from 25 participants, required three components that we expect to be useful beyond this work. First, a synchronization method based on audio tones recorded by both devices; once the drift between the two clocks is modeled, the tone onsets in the two recordings agree to within 0.2~ms. Second, a staged fit of the GNM head model to the tracked landmarks, which separates head pose, eyelid closure and gaze, and the remaining expression. Third, a network that combines a spatial encoder for each electrode grid with a dilated temporal convolutional network, and predicts all 383 expression weights of GNM at 100~Hz. On held-out repetitions of the cued expressions, the median vertex error is 0.43~mm, about half that of a face held at each participant's mean expression. The eye region, which an HMD hides from cameras, is predicted as well as the rest of the face, and even the forehead grid alone retains more than four fifths of the variance of facial motion explained with both grids.

\paragraph*{Limitations and Future Work.}
There are several limitations that we hope to address in future work. Our processing is not causal: the filters are applied forward and backward (Appendix~\ref{sec:emg-processing}), and the network combines 1.27~s of EMG after each frame with 1.27~s before it (Section~\ref{sec:learning}). A prediction is therefore available about 1.3~s after the movement it describes. This delay is acceptable for many interactive applications, but not for real-time animation; a causal filter chain and a causal network would remove it, and we leave this to future work. A related point is that the recording was done in a laboratory setup, with bulky amplifiers and computing; turning this into a wearable system is feasible with current technology and significant engineering effort that is outside the scope of this paper. Each network is trained for one participant, from a session recorded with synchronized video. Decoding the expressions of a new person without such a session remains an open problem, since the electrodes lie over slightly different muscles every time the grids are applied. The network generalizes to new repetitions of the cued expressions, but not yet to free conversation. Including more data, with unscripted speech and expressions in training is the natural next step. Accuracy also depends strongly on the quality of the electrode contact (Section~\ref{sec:performance}), which may get worse using dry electrodes. Our use of high density grids helps, but methods for detecting and compensating for bad contacts should be developed. Finally, a major limitation of the current work is that its ground truth comes from face tracking with a single camera and MediaPipe, which does not capture depth well and has very few landmarks on the forehead. Recent work on multi-view, high-resolution face tracking could improve the quality of the learned expressions (e.g., Bolkart et al.~\cite{Bolkart2026}).

\paragraph*{Prospects.}
Despite these limitations, we have demonstrated a promising new method for facial animation that needs no view of the face, and whose output can be applied directly to other faces and characters. It could have several applications: in animation, to capture an actor's performance without a head-mounted camera rig; in XR, to animate avatars from electrodes built into the face gasket of a headset; in psychology, to measure subconscious emotional responses; and in medicine, to monitor recovery from facial palsy.

\section*{Acknowledgments}
DKP would like to thank Jean-S\'{e}bastien Blouin, Matt Galassi, and Kwang Moo Yi for help with the setup.
Supported in part by NSERC Discovery Grant, CFI, ICICS, and an unrestricted gift from Google.

\bibliographystyle{alpha}
\bibliography{emg2face}

\onecolumn
\begin{center}
  {\LARGE\bfseries Supplementary Materials}
\end{center}
\bigskip

\noindent The supplementary materials contain the following sections:
\begin{itemize}
  \item Appendix~\ref{sec:emg-processing}: \nameref{sec:emg-processing}
  \item Appendix~\ref{sec:interpreting}: \nameref{sec:interpreting}
\end{itemize}

\newpage

\twocolumn
\appendix

\section{EMG Signal Processing}
\label{sec:emg-processing}

\begin{figure*}[!b]
  \centering
  \includegraphics[width=0.88\textwidth]{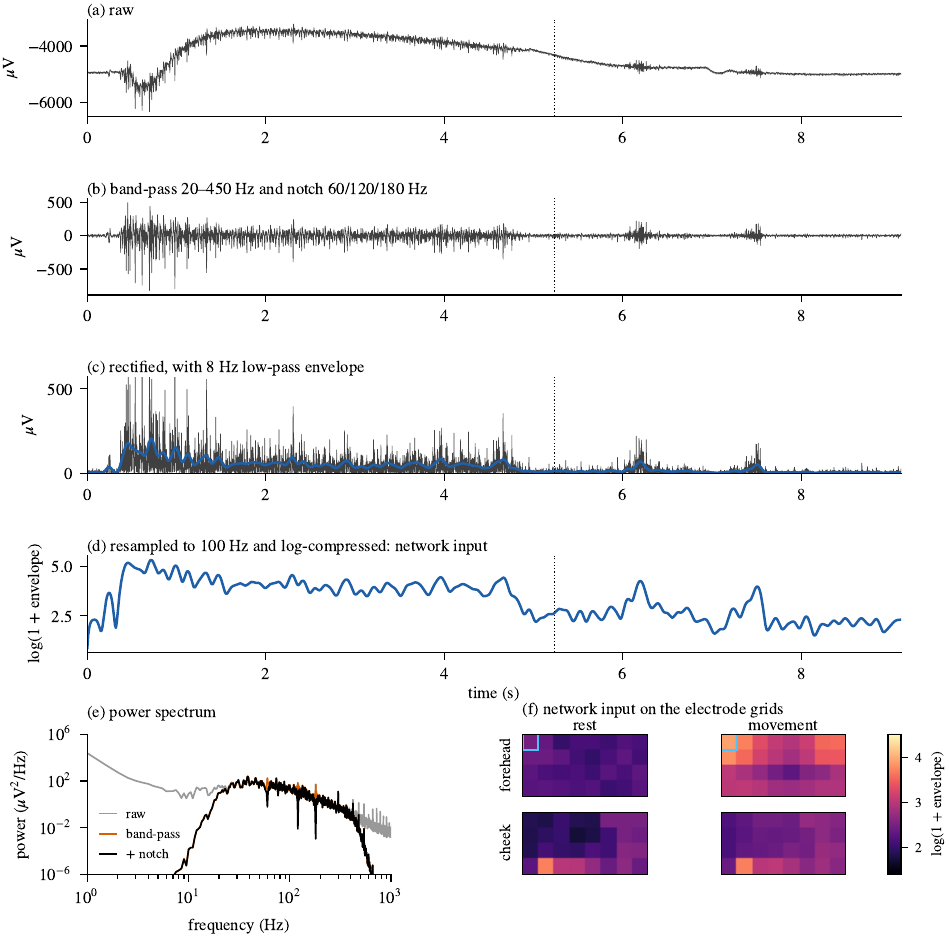}
  \caption{Processing of the raw EMG, for one forehead electrode during a held brow raise. The movement cue is at 0~s and the rest cue at the dotted line. (a) The raw signal is dominated by a large offset and a slow drift. (b) Band-pass and notch filtering isolate the muscle activity. (c) Rectification and low-pass filtering produce the envelope (blue). (d) Resampled to 100~Hz and log-compressed, the envelope is the input to the network. (e) Power spectra of the same electrode: filtering removes the low-frequency drift, which dominates the raw signal, and the mains interference at 60, 120, and 180~Hz. (f) Network input of all 64 electrodes on the forehead and cheek grids, averaged over rest and over the movement; the outlined electrode is the one shown in (a)--(d). Note that the activity rises over the forehead grid, while the cheek grid changes little.}
  \label{fig:emg}
\end{figure*}

The raw EMG is dominated by signals that are not muscle activity. Each channel carries a large offset and a slow drift, of several millivolts in the example of Figure~\ref{fig:emg}a, while the muscle activity itself is a few hundred microvolts. Below 20~Hz, the power of this drift exceeds that of the muscle activity by about three orders of magnitude (Figure~\ref{fig:emg}e). We therefore convert each of the 64 channels into an envelope of muscle activation, in four steps.

First, a band-pass filter (4th-order Butterworth, 20--450~Hz) removes the offset, the drift, and most movement artifacts, which lie below 20~Hz~\cite{DeLuca2010}, together with noise above the bandwidth of surface EMG. Notch filters ($Q = 30$) at 60, 120, and 180~Hz then remove mains interference and its harmonics (Figure~\ref{fig:emg}b). Second, the filtered signal is rectified, and a 4th-order Butterworth low-pass filter at 8~Hz smooths it into an envelope (Figure~\ref{fig:emg}c). Since we process the recordings offline, every filter is applied forward and backward, so that none shifts the envelope in time relative to the video. Third, the envelope is resampled to 100~Hz, by linear interpolation at exact times on the video clock (Section~\ref{sec:sync}). Note that 2048~Hz is not a multiple of 100~Hz, so simply keeping every 20th sample would give 102.4~Hz and accumulate a timing error of about 160~ms over a 7~s trial. The fitted expression weights are interpolated linearly at the same times, and frames in which the face was not tracked are excluded from training and evaluation. Finally, the envelope $e$ is compressed as $\log(1+e)$, which reduces the large difference in level between rest and strong contraction (Figure~\ref{fig:emg}d).

Figure~\ref{fig:emg}f shows the resulting input on both grids, for a held brow raise. The activity rises over the forehead grid, while the cheek grid changes little.

\section{Interpreting the Vertex Error}
\label{sec:interpreting}

The vertex error and $R^2$ of Section~\ref{sec:performance} must be interpreted with caution, for three reasons. First, they are averages over the 5,101 vertices of the visible face and over every frame of the held-out cued trials, while most of the face moves little most of the time. A submillimeter error therefore reflects how little the face moves as much as how well it is predicted, and the error of the mean face is the more meaningful reference. Second, these averages mix regions that move very differently. Third, medians hide failures. We examine the last two in turn.

\paragraph*{Lower face and eye region.}
The lower face moves more than the eye region. Holding each participant's mean face gives a median error of 1.02~mm over the rest of the face, and 0.68~mm over the eye region; the prediction roughly halves both, to 0.50 and 0.36~mm. The lower face, especially the lips and the chin, therefore contributes most of the error of the whole face (a median of 66\%; Figure~\ref{fig:error-map}).

\begin{figure}[ht]
  \centering
  \includegraphics[width=\linewidth]{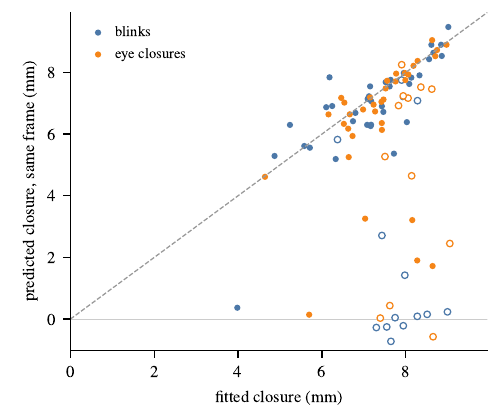}
  \caption{Closure of the eye at the deepest fitted closure of each held-out blink and eye closure, fitted against predicted at the same frame. Trials in which the fitted eye closes by less than 3~mm are omitted, and open circles mark the seven recordings of lower signal quality. Note that most of the blinks that the prediction misses come from these recordings.}
  \label{fig:eyelid-active}
\end{figure}

\paragraph*{The eyelids.}
Most frames hold the eyelids still, so these averages say little about blinks. The palpebral aperture, the distance between the centers of the upper and lower lid margins, measures the lids directly (Figure~\ref{fig:eyelid-active}). At the deepest fitted closure of each held-out blink, the fit closes the eye by a median of 7.5~mm, and the prediction, at the same frame, by 6.7~mm. The median error of the predicted aperture at these frames is 0.39~mm for blinks and 0.20~mm for eye closures. However, medians hide failures. In 11 of 49 blinks, the prediction closes the eye by less than half as much as the fit, or not at all. Ten of these come from the seven recordings of lower signal quality, which contribute only 13 of the blinks. Eye closures are missed in 9 of 47 trials, 4 of them from those recordings. We conclude that blinks are predicted well, except in the recordings of lower signal quality, where the network misses most of them.

\newpage

\end{document}